\documentclass[aip,amsmath,amssymb,reprint,citeautoscript]{revtex4-1}
\usepackage{graphicx}% Include figure files
\usepackage{dcolumn}% Align table columns on decimal point
\usepackage{bm}% bold math
\usepackage[mathlines]{lineno} % Enable numbering of text and display math
\usepackage[utf8]{inputenc}
\usepackage[T1]{fontenc}
\usepackage{mathptmx}
\usepackage{etoolbox}
\usepackage[export]{adjustbox}
\usepackage{xcolor}
\usepackage{xr}
\usepackage{hyperref}
\makeatletter
\def\@email#1#2{%
 \endgroup
 \patchcmd{\titleblock@produce}
  {\frontmatter@RRAPformat}
  {\frontmatter@RRAPformat{\produce@RRAP{*#1\href{mailto:#2}{#2}}}\frontmatter@RRAPformat}
  {}{}
}%
\newcommand*{\addFileDependency}[1]{% argument=file name and extension
\typeout{(#1)}% latexmk will find this if $recorder=0
\@addtofilelist{#1}
\IfFileExists{#1}{}{\typeout{No file #1.}}
}\makeatother

\newcommand*{\myexternaldocument}[1]{%
\externaldocument{#1}%
\addFileDependency{#1.tex}%
\addFileDependency{#1.aux}%
}
\myexternaldocument{Supplementary_Material_for_Higher_Electro-Optic_Response_in_AlScN}

\begin{document}

\preprint{AIP/123-QED}

%\title{Optimizing Electro-Optic Response in AlScN: Ordering, Strain, and Non-polar Superlattice Strategies}

\title{Towards higher electro-optic response in AlScN}

\author{Haochen Wang}
 \affiliation{
 Materials Department, University of California Santa Barbara, Santa Barbara, CA 93106, USA}
\author{Sai Mu}
 \email{mus@mailbox.sc.edu}
 \affiliation{
 Department of Physics and Astronomy, University of South Carolina, Columbia, SC 29208, USA}
\author{Chris G. Van de Walle}
 \email{vandewalle@mrl.ucsb.edu}
 \affiliation{
 Materials Department, University of California Santa Barbara, Santa Barbara, CA 93106, USA}
 
\date{\today}% It is always \today, today,
             %  but any date may be explicitly specified
% insert suggested PACS numbers in braces on next line

\begin{abstract}
Novel materials with large electro-optic (EO) coefficients are essential for developing ultra-compact broadband modulators and enabling effective quantum transduction.
Compared to lithium niobate, the most widely used nonlinear optical material, wurtzite AlScN offers advantages in nano-photonic devices due to its compatibility with integrated circuits.
We perform detailed first-principles calculations to investigate the electro-optic effect in $\mathrm{Al}_{1-x}\mathrm{Sc}_{x}\mathrm{N}$ alloys and superlattices.
At elevated Sc concentrations in alloys, the EO coefficients increase; importantly, we find that cation ordering along the $c$ axis leads to enhanced EO response.
Strain engineering can be used to further manipulate the EO coefficients of AlScN films. 
With applied in-plane strains, the piezoelectric contributions to the EO coefficients increase dramatically, even exceeding 250 pm/V.
We also explore the possibility of EO enhancement through superlattice engineering, finding that nonpolar $a$-plane $\mathrm{(AlN)}_m/\mathrm{(ScN)}_n$ superlattices increase EO coefficients beyond 40 pm/V. 
Our findings provide design principles to enhance the electro-optic effect through alloy engineering and heterostructure architecture.  
\end{abstract}

\maketitle

%The electro-optic (EO) or Pockels effect, which describes the change in the refractive index of a crystal upon application of an electric field, exists only in materials without inversion symmetry.~\cite{pu2010nonlinear}
%The EO effect often coexists with second harmonic generation and the much smaller Kerr effect \cite{boyd2008nonlinear}.
%Here try citing figure S1 in~\ref{fig_asl_piezo}, \ref{fig_3Sc_strain},\ref{fig_asl_el_ion},\ref{fig_mbm}

The Pockels electro-optic (EO) effect describes the linear variation in the refractive index of an optical medium in response to an applied electric field.\cite{pu2010nonlinear}
It occurs only in materials without inversion symmetry,
unlike the (typically much smaller)
%The EO effect often coexists with second harmonic generation 
Kerr effect,\cite{boyd2008nonlinear} which is proportional to the square of the electric field.
Identifying materials with large EO coefficients has garnered increasing attention due to their potential to enable high-speed and low-power EO modulation in integrated photonics applications \cite{reed2010silicon, heck2010hybrid, li2019strong, benner2005exploitation} and quantum transduction in quantum information science. \cite{fan2018superconducting, mckenna2020cryogenic, luo2017chip, chen2019ultra, bruch2019chip}
%Thus, the search for electro-optically active materials is practically important.

In commonly used EO materials such as lithium niobate (LiNbO$_3$, LNO),\cite{abel2013strong, janner2009micro} barium titanate \cite{eltes2019batio, paoletta2021pockels, jiang2020linear} and lead titanate \cite{paillard2019strain} 
%have undergone extensive experimental and theoretical study. 
large EO coefficients primarily arise from the ionic response due to soft optical phonon modes. 
%Also, strain engineering can further increase crystal anharmonicity and soften low-frequency phonon modes, thereby enhancing EO coefficients \cite{hamze2018first, jiang2019designing}.
However, the  difficulty of integrating them with silicon photonics has spurred a search for alternative materials with large EO coefficients.

Aluminum nitride (AlN) has seen increasing use in integrated nano-photonic and opto-mechanical devices due to its excellent compatibility with complementary metal-oxide-semiconductor (CMOS) fabrication. \cite{dong2019aluminum, pernice2012high_apl, pernice2012high_oe, xiong2012low}
Coherent conversion between microwave and optical photons, based on the electro-optic effect, has been successfully accomplished using optical micro-ring cavities made of AlN.\cite{fan2018superconducting}
%However, AlN exhibits lower piezoelectric properties, resulting in reduced sensitivity and electro-mechanical coupling.
However, the EO coefficients of AlN are smaller than those in conventional materials such as LNO (30 pm/V \cite{wooten2000review}).

Alloying offers a promising solution to address this limitation while maintaining CMOS compatibility.\cite{yoshioka2021strongly}
Introducing scandium (Sc) into AlN modifies the crystal structure, enhances its piezoelectric response and can also lead to ferroelectricity. \cite{akiyama2009influence, matloub2013piezoelectric, fichtner2019alscn, wang2021piezoelectric}
Strong piezoelectricity and the presence of ferroelectricity are known to contribute to electro-optic coupling.\cite{wooten2000review,eltes2019batio}
%Therefore, AlScN is expected to significantly increase the electro-optic coefficients.
However, quantitative data on the EO coefficients of AlScN is lacking, and the interplay between piezoelectricity and the EO effect is still to be elucidated.

In this paper, we conduct first-principles calculations to theoretically investigate the EO coefficients in $\mathrm{Al}_{1-x}\mathrm{Sc}_{x}\mathrm{N}$ alloys.
%We systematically analyze both clamped and piezoelectric contributions to the EO coefficients. 
We find that the increase in the EO coefficient $r_{33}$ with Sc concentration ($x_{\mathrm{Sc}}$) is mainly due to  the piezoelectric contribution $r^{\mathrm{piezo}}_{33}$, rather than the clamped term $r^{\mathrm{clamped}}_{33}$.
Based on a systematic analysis of alloy configurations we associate large piezoelectric EO coefficients with the presence of cation ordering along the $c$ axis, which leads to enhanced compliance constants.
We investigate the effect of strain on $\mathrm{Al}_{1-x}\mathrm{Sc}_{x}\mathrm{N}$ alloys, finding that in-plane strains can significantly enhance EO coefficients,
with $r_{33}$ values even exceeding 250~pm/V.
%, exhibiting divergent behaviors near critical strains.
We also explore superlattices, which can enhance control over structural features. 
Nonpolar $a$-plane $\mathrm{(AlN)}_m/\mathrm{(ScN)}_n$ superlattices are particularly promising, showing an increase in both clamped and piezoelectric EO coefficients with Sc layer thickness and reaching $r_{33}$ values up to 40 pm/V.
%, rendering this structure highly promising for applications in nonlinear optics.

All calculations are performed within density functional theory using optimized norm-conserving Vanderbilt pseudopotentials~\cite{hamann2013optimized} and the \texttt{ABINIT} software package.\cite{gonze2005brief}
%(ONCVPSP) generated in PseudoDojo {van2018pseudodojo}
Density functional perturbation theory is employed to calculate the electro-optic coefficients. \cite{veithen2005nonlinear}
The exchange-correlation energy is calculated using the local density approximation (LDA)~\cite{ceperley1980ground, perdew1981self};
more accurate functionals, such as hybrid functionals,\cite{HSE06} are currently not implemented in existing software for electro-optic properties.
The accuracy of using LDA is addressed in Sec. A
%~\ref{sec:scissor} 
of the Supplementary Material (SM).
To simulate the disordered environment in $\mathrm{Al}_{1-x}\mathrm{Sc}_{x}\mathrm{N}$ alloys we use 32-atom supercells which are constructed as $1 \times 2 \times 2$ multiples of the 8-atom orthorhombic unit cell.\cite{supercell_construction}
We also use 48-atom supercells to examine whether the size of the supercell impacts the alloy simulations. We find that using 32- and 48-atom supercells leads to almost identical results.
%with $0 \leq x \leq 0.5$.
We study Sc concentrations from 0 to 50\%; at each concentration
we generate ten inequivalent configurations by randomly placing Sc atoms on cation sites.
A plane-wave cutoff energy of 45 Hartree and a $3 \times 3 \times 2$ Monkhorst-Pack k-point grid are used.\cite{monkhorst1976special}
All structures are fully relaxed until the interatomic forces are smaller than $5 \times 10^{-5}$ Hartree/Bohr.
The physical observables at each Sc concentration are obtained by averaging over the calculated properties of each supercell. 

The electro-optic effect 
%describes how the refractive index ($n_{ij}$) of a crystal changes under an external electric field.
is usually introduced in the context of the optical ellipsoid:\cite{boyd2008nonlinear}
\begin{equation}
    \Delta\left(\frac{1}{n_{ij}^2}\right)=\Delta\left(\varepsilon^{-1}\right)_{i j}=\sum_k r_{i j k} E_k ,
    \label{EO_definition}
\end{equation}
where $n_{ij}$ are the components of the refractive index, $E_k$ the external electric field components, and
$r_{i j k}$ the EO coefficients.

%The EO tensor can be expressed as the sum of three contributions: bare electronic components $r^{\mathrm{elec}}{i j k}$, ionic responses $r^{\mathrm{ion}}{i j k}$, and piezoelectric contributions $r^{\mathrm{piezo}}_{i j k}$ \cite{veithen2005nonlinear}.
The EO tensor can be expressed as the sum of three contributions: bare electronic components $r^{\mathrm{elec}}_{i j k}$, ionic responses $r^{\mathrm{ion}}_{i j k}$ and piezoelectric contributions $r^{\mathrm{piezo}}_{i j k}$.\cite{veithen2005nonlinear}
In the linear response regime, the three contributions are independent and additive ($r_{i j k}^{\mathrm{tot}} = r_{i j k}^{\mathrm{elec}}+r_{i j k}^{\mathrm{ion}} + r_{i j k}^{\mathrm{piezo}}$). 
%In our calculations, the clamped EO coefficients $r^{\mathrm{clamped}}_{i j k}$, which account for electronic and ionic contributions but neglect lattice deformations, are introduced:
The electronic contributions originate from the interaction between the applied electric field and the electron cloud, with ions frozen at their equilibrium positions. The ionic contributions result from $E$-field-induced atomic relaxation while keeping the electron cloud frozen. 
The clamped EO coefficients $r^{\mathrm{clamped}}_{i j k}$ combine these electronic and ionic contributions while neglecting the contributions from lattice deformations, and are defined as follows:
\begin{equation}
    \label{EO_clamped}
    r_{i j k}^{\mathrm{clamped}}=r_{i j k}^{\mathrm{elec}}+r_{i j k}^{\mathrm{ion}}=\frac{-8 \pi}{n_i^2 n_j^2} \chi_{i j k}^{(2)}-\frac{4 \pi}{n_i^2 n_j^2 \sqrt{\Omega_0}} \sum_m \frac{\alpha_{i j}^m l_{m,k}}{\omega_m^2}.
\end{equation}
Here $\chi^{(2)}_{ijk}$ represents the second-order optical susceptibility and $\Omega_0$ denotes the cell volume.
$\alpha_{ij}^m$, the Raman susceptibility of phonon mode $m$, and $l_{m,k}$, the mode polarity\cite{comment_mode_polarity} along the $k$ direction are defined as follows:
\begin{equation}
    \label{raman_sus}
    \alpha_{i j}^m=\sqrt{\Omega_0} \sum_{\kappa, \beta} \frac{\partial \chi_{i j}^{(1)}}{\partial \tau_{\kappa, \beta}} u_m(\kappa \beta),
\end{equation}
\begin{equation}
    \label{mode_polarity}
    l_{m, k}=\sum_{\kappa, \beta} Z_{\kappa, k \beta}^* u_m(\kappa \beta),
\end{equation}
where $u_m(\kappa \beta)$ denotes the phonon eigendisplacement of atom $\kappa$ in the $\beta$ direction of mode $m$.
$\frac{\partial \chi^{(1)}_{ij}}{\partial \tau_{\kappa,\beta}}$ is the derivative of the linear electric susceptibility with respect to atomic displacement, with atom $\kappa$ moving in the $\beta$ direction.
$Z^*_{\kappa,k \beta}$ is the $k$-component of the Born effective charge of atom $\kappa$ moving in the $\beta$ direction.
%The mode polarity $p_{m,k}$ is directly linked to the mode effective charge\cite{gonze1997dynamical},
%\begin{equation}
%    \label{mode_effective_charge}
%    Z_{m, \alpha}^*=\frac{\sum_{\kappa \beta} Z_{\kappa, \alpha \beta}^* U_{m \mathbf{q}=\mathbf{0}}(\kappa \beta)}{\left(\sum_{\kappa \beta}\left[U_{m \mathbf{q}=\mathbf{0}}(\kappa \beta)\right]^* U_{m \mathbf{q}=0}(\kappa \beta)\right)^{1 / 2}} .
%\end{equation}

%The electronic component originates from the interaction between the external field and valence electrons, with ions artificially clamped at their equilibrium positions.The ionic contribution to the EO tensor results from atomic relaxation under the applied electric field ${E}k$ and from variations in the dielectric constants $\epsilon{ij}$ induced by these displacements.The piezoelectric contribution arises from lattice shape relaxation due to the converse piezoelectric effect. 
%In linear response region, the three contributions to the EO coefficients are independent and additive. 

%Most quantities in Eq.\,\ref{EO_clamped} are calculated using density function perturbation theory.\cite{veithen2005nonlinear}  

%The piezoelectric contribution arises from the change in dielectric constant associated with the piezoelectric effect;
The piezoelectric contribution arises from the lattice deformation induced by the electric field, which in turn changes the dielectric constants;
it can be computed using the piezoelectric strain coefficients $d_{k \mu \nu}$ and the elasto-optic coefficients $p_{i j \mu \nu}$ (which represent the derivative of the inverse dielectric tensor with respect to the strain):
\begin{equation}
    \label{EO_piezo}
    r_{i j k}^{\text {piezo }}=\sum_{\mu, \nu=1}^3 p_{i j \mu \nu} d_{k \mu \nu}.
\end{equation}

In the subsequent discussion, Voigt notation is adopted. 
Due to the symmetry of indices $i$ and $j$, $r_{ijk}$ can be contracted to a $6 \times 3$ tensor $r_{hk}$.
Similarly, piezoelectric strain coefficients $d_{k \mu \nu}$ can be represented by a $6 \times 3$ tensor and  elasto-optic coefficients $p_{i j \mu \nu}$  by a $6 \times 6$ tensor.  
Which components of the tenors are independent and nonzero can be determined by a symmetry analysis of the crystal.

%In linear response region, the three contributions to the EO coefficients are independent and additive. 
%As previously described, we model the $\mathrm{Al}_{1-x}\mathrm{Sc}_{x}\mathrm{N}$ alloys using 32-atom supercells, as illustrated in Fig. \ref{fig_EO_alloy}(a).
%As previously described, we model the $\mathrm{Al}_{1-x}\mathrm{Sc}_{x}\mathrm{N}$ alloys using multiple 32-atom supercells.
Our actual results at each Sc concentration are obtained by averaging over ten different supercells with randomly placed cations.
%therein and average the calculated physical properties over these supercells.
Calculated lattice parameters $a$, the $c/a$ ratio, and the internal displacement parameter $u$ are plotted in %Fig.~\ref{fig_lattice}
Fig. S1
 %(Sec.~\ref{sec:lattice_geometry} in the SM).
 (Sec. B in the SM).
They compare well with experiments. \cite{Akiyama2009,Schonweger2022}
The phase transition from wurtzite to a metastable layered-hexagonal structure (with $u$ = 0.5) occurs between 37.5\% and 43.75\%, consistent with previous findings.\cite{wang2021piezoelectric,Akiyama2009,Satoh2022_JAP}

\begin{figure}
    \includegraphics[width=0.45\textwidth]{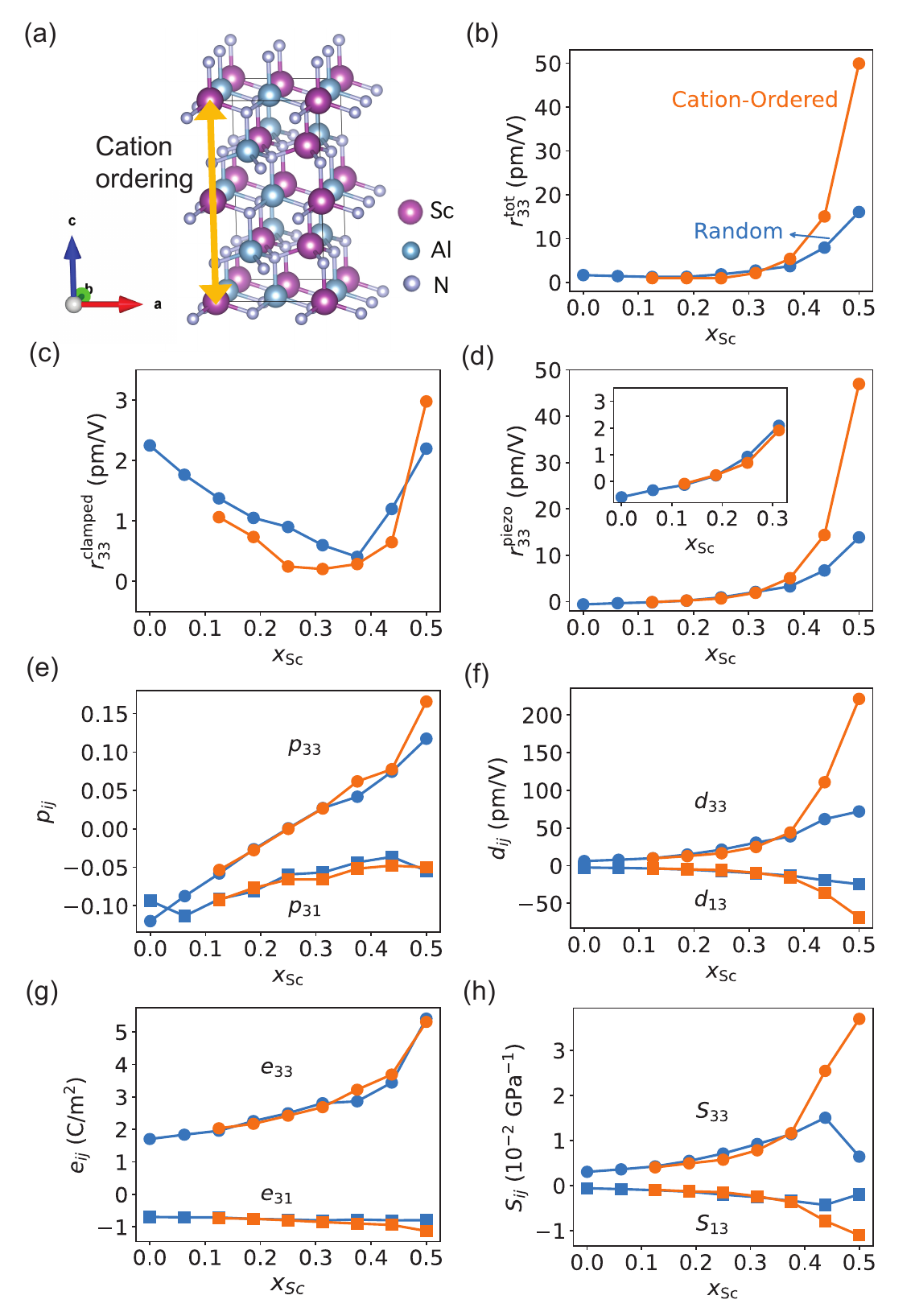}
    \caption{(a) Atomic structure of a 32-atom supercell of $\mathrm{Al_{0.5}Sc_{0.5}N}$. The orange arrow denotes a third-nearest cation-cation neighbor. (b) Total, (c) Clamped and (d) Piezoelectric EO coefficients $r_{33}$ (the inset shows $r_{33}^\mathrm{piezo}$ at $x_{\mathrm{Sc}} \leq 31.25\%$); (e) Elasto-optic coefficients $p_{33}$ and $p_{31}$; (f) Piezoelectric strain coefficients $d_{33}$ and $d_{13}$; (g) Piezoelectric stress coefficients $e_{33}$ and $e_{31}$; (h) Compliance constants $S_{33}$ and $S_{13}$. Blue symbols represent averaged values over randomly generated alloy configurations, while orange symbols represent averages over structures with cation ordering along the $c$ axis.}
    \label{fig_EO_alloy}
\end{figure}

The wurtzite structure has the $6mm$ point group, giving three independent nonzero elements in the EO tensor: $r_{13}$, $r_{33}$, and $r_{51}$. 
Among these, $r_{33}$ (=$r_{333}$=$r_{zzz}$), which quantifies the response of the $zz$ component of the inverse dielectric tensor to the applied field along $z$, is the largest.
A fully disordered solid solution should retain the wurtzite symmetry.
However, the finite-size supercells that we use to model the alloy do lower the symmetry, thus introducing additional nonzero elements such as $r_{11}$ and $r_{22}$. 
Our calculations indicate that 
%$r_{33}$ and $r_{13}$ remain dominant (Fig.~\ref{fig_EO_alloy} and Fig.~\ref{fig_r13} in SM), and 
such additional components are small and close to zero when averaging over multiple supercells (see %Sec.~\ref{sec:r_11_22} in the SM).
Sec. C in the SM).

%As depicted in the inset of Fig. \ref{fig_EO_alloy}(b), $r_{33}^{\mathrm{clamped}}$ initially decreases and then increases past the turning point at the phase transition concentration. Conversely, $r_{33}^{\mathrm{piezo}}$ monotonically increases across the entire range of scandium concentration, significantly contributing to the total $r_{33}$, particularly at high concentrations.

The results of our systematic investigation of the EO effect are shown in Fig.~\ref{fig_EO_alloy}.
%, for the example of a $\mathrm{Al_{0.5}Sc_{0.5}N}$ alloy. 
For alloy configurations with randomly placed cations (labeled ``Random''), $r_{33}^{\mathrm{clamped}}$ first decreases with $x_{\mathrm{Sc}}$ and then increases, with a turning point roughly corresponding to the phase-transition concentration. 
In contrast, $r_{33}^{\mathrm{piezo}}$ monotonically increases over the entire range of Sc concentrations, starting out smaller than the clamped contribution at low concentrations but significantly contributing to the total $r_{33}$ at concentrations above 30\%. 
%It is crucial to note that at high Sc concentrations (43.75\% and 50\%), the variability of both clamped and piezoelectric EO coefficients across ten random configurations is significant, due to structural fluctuations. In this case, identifying the most contributing structures among the generated configurations is essential. 

The variations of both clamped and piezoelectric EO coefficients among different configurations of random alloys are small at low $x_{\mathrm{Sc}}$ but become significant at high $x_{\mathrm{Sc}}$ 
%(see Sec.~\ref{sec:variations_r33} in the SM). 
(see Sec. D in the SM). 
%For the purpose of designing high-EO materials, identifying the structural features of supercells with high EO (at high Sc concentrations) is pivotal. 
We now focus on identifying the structural features of alloy configurations that give rise to particularly high EO coefficients, with the goal of providing guidelines for design or synthesis of high-EO materials.
Through analyses of pair distribution functions and ruling out the possibility that \textit{pairs} of cations cause enhancements, we found that a common structural feature gives large EO coefficients, namely the formation of same-species cation chains along the $c$ axis [see Fig.~\ref{fig_EO_alloy}(a)].  
While we do not aim to address the details of the alloy energetics, we observe that at every concentration the structure with the lowest energy is invariably a cation-ordered structure, rendering their occurrence in experimental samples quite likely.
%This corresponds to a long-range ordering along the $c$ axis (``$c$-ordering'').
To confirm the correlation of this ordering with enhancements in the EO coefficients, we generate multiple additional cation-ordered configurations at each concentration, deliberately forming same-species chains along the $c$ axis.

The following analysis focuses on $r_{33}$, but similar results are obtained for $r_{13}$.
In Fig.~\ref{fig_EO_alloy}(b), we plot averaged EO coefficients $r_{33}^{\mathrm{tot}}$ of random alloys (blue) and cation-ordered structures (orange). 
The $r_{33}^{\mathrm{tot}}$ of cation-ordered structures
%monotonically increases and 
is significantly more enhanced at high $x_{\mathrm{Sc}}$ compared to the random structures.
Comparing Fig.~\ref{fig_EO_alloy}(c) and (d), this is caused by the enhancement of piezoelectric EO contributions in cation-ordered structures.
%(So is $r_{13}^{\mathrm{piezo}}$, see Fig.~\ref{fig_r13} (a)).
Using Voigt notation and taking advantage of the wurtzite symmetry in Eq.~(\ref{EO_piezo}), $r^{\mathrm{piezo}}_{33}$ can be written as $r^{\mathrm{piezo}}_{33} = 2p_{31}d_{13} + p_{33}d_{33}$.
Figures~\ref{fig_EO_alloy}(e) and (f) illustrate the piezoelectric strain coefficients ($d_{ij}$) and the elasto-optic coefficients ($p_{ij}$) for both random and cation-ordered structures. We find that
(1) $p_{33}$ and $d_{33}$, increasing faster than $p_{31}$ and $d_{13}$, dominate the contributions to $r^{\mathrm{piezo}}_{33}$; 
(2) $p_{33}$ and $p_{31}$ are relatively insensitive to the details of the structure, given similar results of random and cation-ordered structures;
and (3) $d_{33}$ and $d_{13}$ are not sensitive to structure up to $x_{\mathrm{Sc}}=37.5\%$, but are significantly enhanced in the cation-ordered structures at 43.75\% and 50\%.
We can therefore attribute the enhancement of $r^{\mathrm{piezo}}_{33}$ in cation-ordered structures at high $x_{\mathrm{Sc}}$ to the rise in piezoelectric strain coefficients $d_{33}$ and $d_{13}$.

We explore the enhancement in more detail by writing the piezoelectric strain coefficients $d_{i j}$ as a product of the piezoelectric stress tensor $e_{i k}$ and the compliance tensor $S_{k j}$:
\begin{equation}
    \label{compliance_tensor}
    d_{i j} = \sum_{k}{e_{i k} S_{k j}}.
\end{equation}
Figure~\ref{fig_EO_alloy}(g) and (h) show that $e_{33}$ and $e_{31}$ have similar values in random and cation-ordered structures, but $S_{33}$ and $S_{13}$ show significant enhancements in cation-ordered structures at Sc concentrations of 43.75\% and 50\%.
The larger $S_{33}$ values in the cation-ordered structures reflect weaker cation-nitrogen bond strength along the $c$ axis, as confirmed by a crystal orbital Hamilton population analysis~\cite{COHP,LOBSTER} 
%(see Sec.~\ref{sec:ICOHP_analysis} in the SM).
(see Sec. E in the SM).

%From another perspective, both ``high Sc content'' and ``cation ordering'' are essential to enhance the EO response in $\mathrm{Al}_{1-x}\mathrm{Sc}_x\mathrm{N}$ alloys, as they push the system closer to the phase transition.
%When only ``high Sc content'' is present, the random alloys transition from the wurtzite phase to a metastable layered-hexagonal phase (Fig.~\ref{fig_lattice}). 
%Cation ordering along the $c$ axis reduces both the in-plane lattice parameter $a$ and the $u$ parameter, thus driving the system back to the wurtzite phase and closer to the phase transition.
%When cation ordering is present but at low Sc concentrations, the alloys remain in the wurtzite phase. In this scenario, raising Sc concentration increases the in-plane lattice parameter $a$ and $u$-parameter, pushing the wurtzite system towards a metastable layered-hexagonal phase and closer to the phase transition.
%In addition to the cation ordering and Sc concentration, applying in-plane strains externally can also modify the lattice parameters, further driving the system towards the phase transition.
%%We conclude that cation-cation ordering  significantly enhances EO responses beyond 37.5\% Sc through the enhanced compliance constants $S_{33}$.

The combination of high Sc content and cation ordering enhances the EO response by pushing the system closer to the phase transition.
When we increase the Sc content of random alloys, the in-plane lattice parameter $a$ and the $u$ parameter increase, but the alloys transition from the wurtzite phase to a layered-hexagonal phase (Fig. S1).   %(Fig.~\ref{fig_lattice}). 
Cation ordering along the $c$ axis counteracts the effect of rising Sc concentration, reducing both the in-plane lattice parameter $a$ and the $u$ parameter, thus driving the system back to the wurtzite phase and closer to the phase transition.

\begin{figure}
    \centering
    \includegraphics[width=0.45\textwidth]{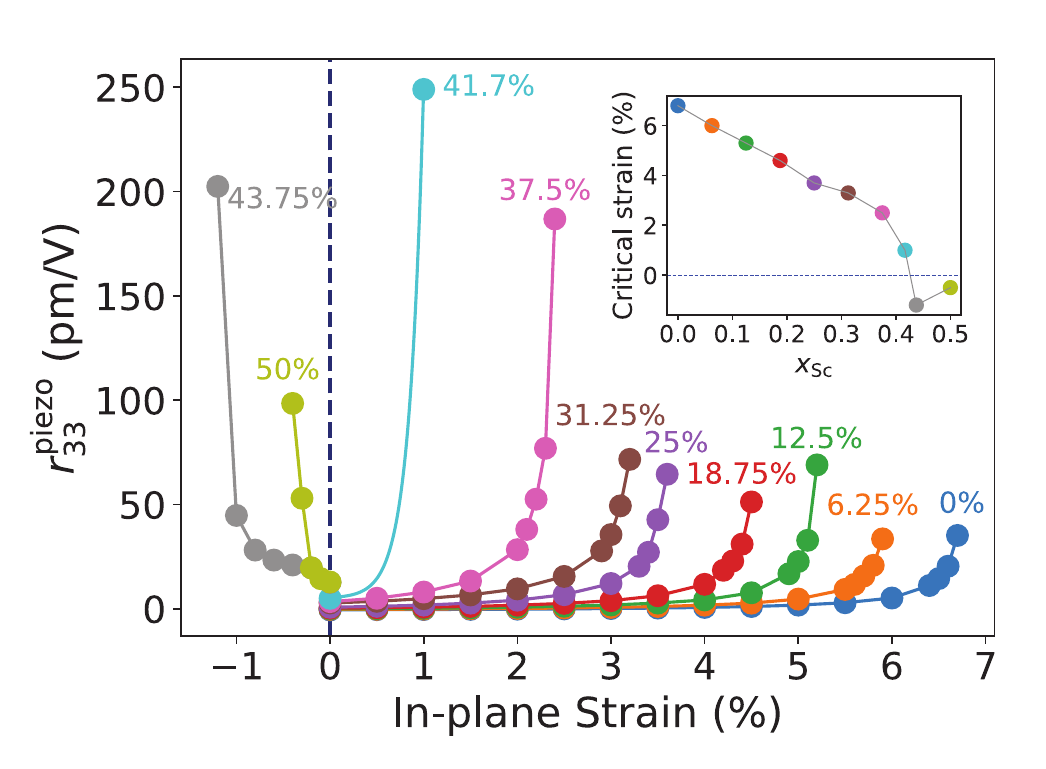}
    \caption{Piezoelectric EO coefficients $r_{33}^{\mathrm{piezo}}$ of $\mathrm{Al}_{1-x} \mathrm{Sc}_x \mathrm{N}$ as a function of in-plane strain (positive for tensile strain, negative for compressive strain). The inset shows the decrease in critical strain \cite{critical_strain} values as a function of Sc concentration $x_{\mathrm{Sc}}$.}
    \label{fig_EO_strain}
\end{figure}

%Inspired by the significant EO coefficients induced by strain in $\mathrm{SrTiO_3}$\cite{hamze2018first}, we investigate the effects of strain engineering on $\mathrm{Al_{1-x} Sc_x N}$ alloys.  
These insights about the role of Sc concentration and cation ordering suggest that manipulating the lattice parameter by applying in-plane strain could also be fruitful.
%Next we investigate the impact of strain on EO coefficients.
Due to the mismatch of in-plane lattice parameters, strain is likely to be present in heteroepitaxially grown films, such as $\mathrm{Al}_{1-x} \mathrm{Sc}_x \mathrm{N}$ grown on GaN~\cite{2022Casamento_Epitaxial} or Si~\cite{2020Park_IEEE} substrates.
%Given that strain may have an impact on the piezoelectric constants and EO coefficients, here we explicitly investigate the epitaxial strain dependence of EO in $\mathrm{Al}_{1-x} \mathrm{Sc}_x \mathrm{N}$. 
Large strain effects on EO coefficients were experimentally and computationally demonstrated in the cases of $\mathrm{SrTiO_3}$\cite{hamze2018first,2019APL_Large}, strained silicon\cite{2011Bartos_strainedsilicon,2006Nature_Jacobsen} and $c$-plane $\mathrm{(AlN)}_1/\mathrm{(ScN)}_1$ superlattices.\cite{2019Jiang_PRL}

At each Sc concentration, we select one configuration of the $\mathrm{Al}_{1-x} \mathrm{Sc}_x \mathrm{N}$ random alloy and apply in-plane strain by constraining the in-plane lattice parameters and relaxing the out-of-plane lattice parameter and all internal degrees of freedom.  
%In Sec.~\ref{sec:independence_strain} of the SM we demonstrate that our conclusions regarding strain do not depend on the specific choice of alloy configuration.
In Sec. F of the SM we demonstrate that our conclusions regarding strain do not depend on the specific choice of alloy configuration.
The results in Fig.~\ref{fig_EO_strain} show that
for $x_{\mathrm{Sc}}$ below 43.75\%, $r^{\mathrm{piezo}}_{33}$ first increases slowly with in-plane \textit{tensile} strains, then dramatically rises as the strain approaches a critical value. 
However, at higher $x_{\mathrm{Sc}}$ (43.75\% and 50\%), in-plane \textit{compressive} strain is needed to enhance $r^{\mathrm{piezo}}_{33}$.
%and the $r^{\mathrm{piezo}}_{33}$ of $\mathrm{Al_{0.5} Sc_{0.5} N}$ exceeds 200 pm/V. 
%This difference is attributed to the phase transition 
%of $\mathrm{Al}_{1-x} \mathrm{Sc}_x \mathrm{N}$.
%from wurtzite to layered-hexagonal, which
%happens between 37.5\% and 43.75\% and which is accompanied by an increase in the in-plane lattice parameter $a$ (see Fig.~\ref{fig_lattice} in the SM).
%The alloys maintain wurtzite structures up to this regime but transition to layered-hexagonal structures beyond that.

The difference in the sign of the strain needed for enhancement makes clear that the strain has an impact because it drives the system closer to the phase transition:
for wurtzite  ($x_{\mathrm{Sc}} \leq 37.5\%$), in-plane \textit{tensile} strain is required to increase $u$ and drive the system towards a layered-hexagonal phase, while for the metastable layered-hexagonal structures ($x \ge 43.75\%$), we need in-plane \textit{compressive} strain to decrease $u$ and push the system back to wurtzite.
%can both drive the system towards the phase transition regime which will boost the EO response.
%In-plane compressive strains on layered-hexagonal structures decrease in-plane lattice constants (increase the out-of-plane lattice constant $c$), and further decrease the internal parameter $u$, nudging the layered-hexagonal structures back towards wurtzite structures.

%In Sec.~\ref{sec:decomposition_r33} of the SM, we analyze the $r^{\mathrm{clamped}}_{33}$ and all factors contributing to $r^{\mathrm{piezo}}_{33}$ under in-plane strains. 
In Sec. G of the SM, we analyze the $r^{\mathrm{clamped}}_{33}$ and all factors contributing to $r^{\mathrm{piezo}}_{33}$ under in-plane strains. 
%Figure~\ref{fig_strain_engi} shows that $r^{\mathrm{clamped}}_{33}$, $p_{33}$ and $e_{33}$ increase monotonically and almost linearly with strain at $x_{\mathrm{Sc}}$ below 43.75\%.
Figure S6 shows that $r^{\mathrm{clamped}}_{33}$, $p_{33}$ and $e_{33}$ increase monotonically and almost linearly with strain at $x_{\mathrm{Sc}}$ below 43.75\%.
For $r^{\mathrm{piezo}}_{33}$, its dramatic enhancement near the critical strain mainly originates from the piezoelectric strain coefficients $d_{33}$ and the compliance constants $S_{33}$. %similar to Fig. \ref{fig_EO_alloy}.

The inset of Fig.~\ref{fig_EO_strain} shows that the critical strain decreases as a function of $x_{\mathrm{Sc}}$. 
This means that a relatively small strain can greatly enhance the EO response in $\mathrm{Al}_{1-x} \mathrm{Sc}_x \mathrm{N}$ alloys with high Sc content. 
As a test, we modeled a 41.7\% alloy (in a 48-atom supercell) and found that a 1\% strain suffices to increase $r_{33}$ to 251~pm/V.
%At $x_{\mathrm{Sc}}=41.7\%$ (presented by 48-atom supercells $\mathrm{Al}_{14} \mathrm{Sc}_{10} \mathrm{N}_{24}$), we observed great EO response ($r_{33}=251$ pm/V) at a small critical strain of 1\%.
Achieving strains large enough to make a significant difference at low $x_{\mathrm{Sc}}$ may be challenging.
We note however, that inhomogeneities in the alloy could lead to locally enhanced Sc concentrations, and potentially local strains as well. 
Since the strain enhancement is highly nonlinear, it is thus conceivable that significant increases in EO coefficients could be observed in inhomogeneous alloys with nominal Sc concentrations well below the phase transition.
%To investigate this effect, a specific configuration of $\mathrm{Al_{0.75} Sc_{0.25} N}$ was constructed where Sc atoms are aggregated, generating 8\% local strain near the Sc-rich region. 
%Local strains were estimated to be as high as 8\% by comparing the geometry adjacent to the Sc-aggregated region. 
%This supports the potential for observing strong EO effect even at lower Sc concentrations, through the enlarged local strain in Sc-segregated samples.

\begin{figure}
    \centering
    \includegraphics[scale=0.6]{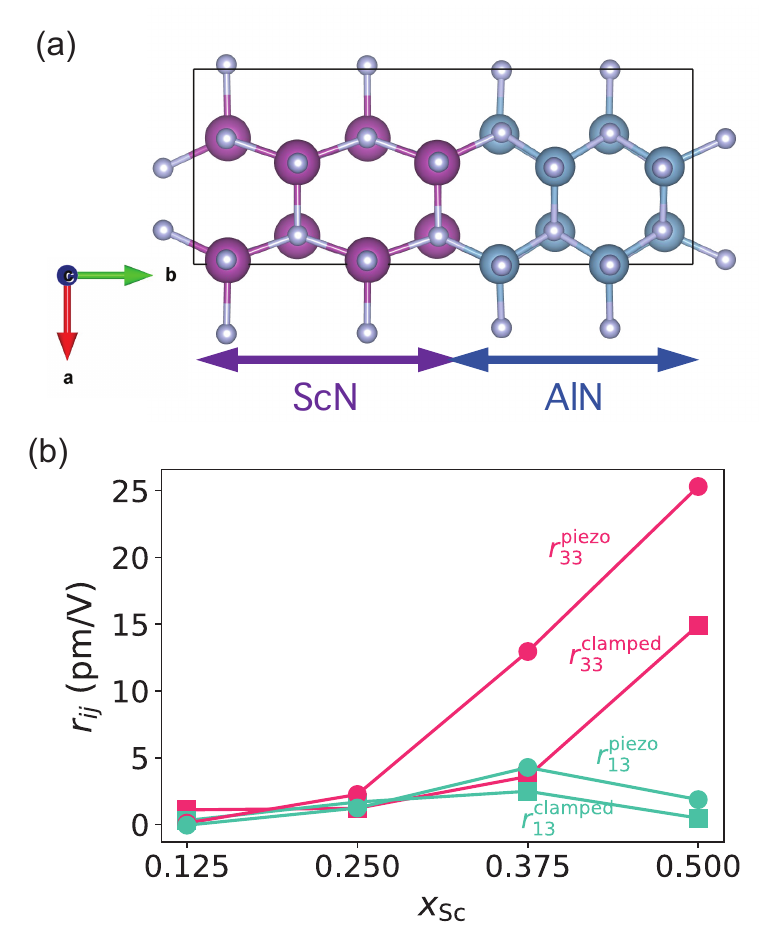}
    \caption{(a) Atomic structure of an $a$-plane  $\mathrm{(AlN)}_m/\mathrm{(ScN)}_n$ superlattice with $m=n=4$ and thus $x_{\mathrm{Sc}}=50\%$. (b) Clamped and piezoelectric contributions to $r_{33}$ and $r_{13}$ for $x_{\rm Sc}=n/(m+n)$.
    %In $\mathrm{(AlN)_m/(ScN)_n}$ superlattices, the Sc concentration means the ratio of $\frac{n}{m+n}$. 
    %We keep the number of atomic layers along b axis constant ($m+n=8$ as (a) shows) and change $n$ from 1 to 4.
    }
    \label{fig_EO_superlattice}
\end{figure}

Our finding that the cation ordering along the $c$ axis can magnify the EO coefficients inspires us to design structures that would build in such order.
Superlattices offer this potential because they can be grown with atomic precision.
Since the relevant cation ordering is along the $c$ axis, we focus on structures in which the $c$ axis lies in the growth plane, which is the case for non-polar $a$-plane superlattices [Fig.~\ref{fig_EO_superlattice}(a)].\cite{c_plane_SL}
%Inspired by this observation, we can design a superlattice which magnifies the short range order between the third nearest cation interactions and explore its EO response. 
%This is achieved by constructing the a-plane $\mathrm{(AlN)_m/(ScN)_n}$ superlattice (see Fig.\,\ref{fig_EO_superlattice}(a) for structural illustration). 
%The a-plane of a wurtzite structure is non-polar and should be distinguished from the common polar c-plane (001). 
We perform systematic studies of $a$-plane $\mathrm{(AlN)}_m/\mathrm{(ScN)}_n$ superlattices in which the total number of atomic layers along the $b$-axis remains constant ($m+n=8$) and vary the number of ScN layers $n$ from 1 to 4, resulting in Sc concentrations of 12.5\%, 25\%, 37.5\%, and 50\%.

The resulting clamped and piezoelectric contributions to $r_{13}$ and $r_{33}$ are plotted in Fig.~\ref{fig_EO_superlattice}(b).
%The curves for both coefficients are closely aligned at concentrations below 25%.  
%The clamped and piezoelectric contributions to $r_{13}$ and $r_{33}$ are comparable up to 25\% Sc, beyond that, both contributions to $r_{33}$ increase rapidly while those to $r_{13}$ remain inert to Sc content.
$r_{13}$ is relatively insensitive to Sc content, but both contributions to
$r_{33}$ rapidly increase above $x_{\mathrm{Sc}}=25\%$.
At $m=n=4$, i.e., $x_{\mathrm{Sc}}=50\%$ , the total $r_{33}$ reaches up to 40 pm/V,
%(15 pm/V from $r_{33}^{\mathrm{clamped}}$ and 25 pm/V from $r_{33}^{\mathrm{piezo}}$), 
representing a twenty-fold enhancement over pure AlN and surpassing the EO coefficient of LNO. 
%In Sec.~\ref{sec:smaller_superlattice} of the SM, we also study $a$-plane superlattices with $m=n=1, 2$ and 3 at $x_{\mathrm{Sc}}=50\%$, and similarly observe enhancements of the EO coefficients.
In Sec. H of the SM, we also study $a$-plane superlattices with $m=n=1, 2$ and 3 at $x_{\mathrm{Sc}}=50\%$, and similarly observe enhancements of the EO coefficients.

%The reason for the increase in $r^{\mathrm{piezo}}_{33}$ is the same as in alloys, namely a substantial piezoelectric coefficient $d_{33}$ combined with a large compliance constant $S_{33}$ (see Fig.~\ref{fig_asl_piezo}).
The reason for the increase in $r^{\mathrm{piezo}}_{33}$ is the same as in alloys, namely a substantial piezoelectric coefficient $d_{33}$ combined with a large compliance constant $S_{33}$ (see Fig. S8).
Interestingly, Fig.~\ref{fig_EO_superlattice}(b) shows that in the superlattice structures not just $r_{33}^{\mathrm{piezo}}$ reaches large values at high Sc content, but also $r_{33}^{\mathrm{clamped}}$.  This contrasts with the behavior in $\mathrm{Al}_{1-x} \mathrm{Sc}_x \mathrm{N}$ alloys, where $r_{33}^{\mathrm{clamped}}$ always remain modest.

%To investigate the origin of the enhancement in $r_{33}^{\mathrm{clamped}}$, we compare $r^{\mathrm{ion}}_{33}$ with $r^{\mathrm{elec}}_{33}$ in Fig.~\ref{fig_asl_el_ion} (Sec.~\ref{sec:superlattice} of the SM) at $x_{\mathrm{Sc}}=50\%$, showing that the former dominates $r_{33}^{\mathrm{clamped}}$. 
To investigate the origin of the enhancement in $r_{33}^{\mathrm{clamped}}$, we compare $r^{\mathrm{ion}}_{33}$ with $r^{\mathrm{elec}}_{33}$ in Fig. S9 (Sec. I of the SM) at $x_{\mathrm{Sc}}=50\%$, showing that the former dominates $r_{33}^{\mathrm{clamped}}$. 
%Sec.~\ref{sec:superlattice} of the SM shows that the ionic response $r^{\mathrm{ion}}_{33}$ dominates $r_{33}^{\mathrm{clamped}}$ at $x_{\mathrm{Sc}}=50\%$.
According to Eq.~(\ref{EO_clamped}), we decompose $r^{\mathrm{ion}}_{33}$ mode-by-mode into the Raman susceptibility $\alpha^m_{33}$ and the mode polarity $l_{m,3}$.
%For 32-atom superlattices, 96 phonon modes (3 acoustic and 93 optical modes) are analyzed in Fig.~\ref{fig_mbm}.
For 32-atom superlattices, 96 phonon modes (3 acoustic and 93 optical modes) are analyzed in Fig. S10.
At 50\%, two phonon modes, in which AlN and ScN layers moving towards opposite directions along $c$, contribute most strongly to the EO response.
These two shearing modes have large Raman susceptibility $\alpha^m_{33}$ and mode polarity $l_{m,3}$ at low frequencies (139 and 254 cm$^{-1}$),  leading to the substantial ionic EO contributions to $r_{33}^{\mathrm{clamped}}$.

%All our calculations employ the LDA functional which underestimates the band gap of materials, potentially affecting the electro-optic properties.
%The use of more accurate functionals, e.g., hybrid functinals, \cite{HSE06} is currently not implemented in existing software for electro-optic properties.
%To assess the potential impact of using LDA on the EO coefficients, we apply a scissor correction in two cases, AlN and $\mathrm{Al_{0.5}Sc_{0.5}N}$, manually adjusting the band gap to the experimental value.  
%Adjusting the band gap changes the total EO coefficients by less than  17\%, confirming the reliability of the LDA calculations.
%More details are provided in the Sec.~\ref{sec:scissor} of the SM.

In summary, we have reported first-principles investigations of the EO effect in $\mathrm{Al}_{1-x} \mathrm{Sc}_x \mathrm{N}$ alloys and $a$-plane $\mathrm{(AlN)}_m / \mathrm{(ScN)}_n$ superlattices.
For alloys, we find that the piezoelectric contributions $r^{\mathrm{piezo}}_{33}$ 
%are comparable to 
dominate over $r^{\mathrm{clamped}}_{33}$ at $x_{\mathrm{Sc}}$ above 30\%.
We identify that $r^{\mathrm{piezo}}_{33}$ is enhanced by cation ordering that weakens the cation-nitrogen bonds along the $c$ axis, resulting in large compliance constants $S_{33}$.
We find that in-plane strains can dramatically increase $r^{\mathrm{piezo}}_{33}$ by driving the system closer to the phase transition.
Non-polar $a$-plane $\mathrm{(AlN)}_m/\mathrm{(ScN)}_n$ superlattices can also be used to significantly enhance both clamped and piezoelectric contributions to $r_{33}$.
%Similarly, the piezoelectric EO coefficients are linked to mechanical properties.
%Through a mode-by-mode decomposition, it was found that Raman susceptibilities $\alpha_{33}^m$ of low-frequency phonon modes play a decisive role in influencing $r^{\mathrm{ion}}_{33}$.
The insights provided by our work provide guidelines for synthesizing $\mathrm{Al}_{1-x} \mathrm{Sc}_x \mathrm{N}$ alloys and heterostructures with optimized electro-optic coefficients.
\\

See the Supplementary Material for the sensitivity of EO coefficients to the choice of functional, lattice geometries, the symmetry of EO tensor and variations of $r_{33}$ in $\mathrm{Al}_{1-x} \mathrm{Sc}_x \mathrm{N}$ alloys, bond strength evaluation of random and cation-ordered structures, independence of conclusions for strained structures on alloy configuration, decomposition of $r_{33}$ in the presence of strain, and the EO coefficients of $a$-plane $\mathrm{(AlN)}_m/\mathrm{(ScN)}_n$ superlattices.

\begin{acknowledgments}
    The authors appreciate fruitful discussions with G. Yang and Profs.~C. Dreyer, H. Tang, and D. Jena. 
    This work is supported by the Army Research Office (W911NF-22-1-0139) and by SUPREME, one of seven centres in JUMP 2.0, a Semiconductor Research Corporation program sponsored by the Defense Advanced Research Projects Agency. 
%SUPeRior Energy-efficient Materials and dEvices (SUPREME) under the grant No. XXX. 
    This work used Bridges-2 at Pittsburgh Supercomputing Center (PSC) through allocation DMR070069 from the Advanced Cyberinfrastructure Coordination Ecosystem: Services \& Support (ACCESS) program, which is supported by National Science Foundation grant No. 2138259, 2138286, 2138307, 2137603 and 2138296. This research also used resources of the National Energy Research Scientific Computing Center, a DOE Office of Science User Facility supported by the Office of Science of the U.S. Department of Energy under Contract No. DE-AC02-05CH11231 using NERSC award BES-ERCAP0028497.
\end{acknowledgments}

\section*{AUTHOR DECLARATIONS}
\subsection*{Conflict of Interest}
The authors have no conflicts to disclose.

\section*{Data Availability}
The data that support the findings of this study are available from the corresponding author upon reasonable request.

%\nocite{*}
\bibliography{EO_reference}% Produces the bibliography via BibTeX.

\begin{thebibliography}{50}
\bibitem [{sup()}]{supercell_construction}%
  \BibitemOpen
  \href@noop {} {}\bibinfo {note} {The lattice vectors in the primitive cell of
  wurtzite AlN are $\mathbf{a} = a_0 (\frac{\sqrt{3}}{2},\frac{1}{2},0)$,
  $\mathbf{b} = a_0 (-\frac{\sqrt{3}}{2},\frac{1}{2},0)$ and $\mathbf{c} = c_0
  (0, 0, 1)$. An orthorhombic 8-atom conventional unit cell is constructed with
  lattice vectors $\mathbf{a'}=\mathbf{a}+\mathbf{b}$,
  $\mathbf{b'}=\mathbf{a}-\mathbf{b}$ and
  $\mathbf{c'}=\mathbf{c}$.}\BibitemShut {Stop}%
\bibitem [{com()}]{comment_mode_polarity}%
  \BibitemOpen
  \href@noop {} {}\bibinfo {note} {We use the symbol $l_{m,k}$ for mode
  polarity, rather than the more customary $p_{m,k}$, to avoid confusion with
  the elasto-optic coefficients.}\BibitemShut {Stop}%
\bibitem [{c_p()}]{c_plane_SL}%
  \BibitemOpen
  \href@noop {} {}\bibinfo {note} {For comparison, we also investigated
  $c$-plane superlattices (which do not have the ordering along the $c$ axis);
  we observed no enhancements of EO coefficients.}\BibitemShut {Stop}%
\end{thebibliography}

\end{document}

%\preprint{AIP/123-QED}

\title{Supplementary Material for ``Towards higher electro-optic response in AlScN''}

\author{Haochen Wang}
 \affiliation{
 Materials Department, University of California Santa Barbara, Santa Barbara, CA 93106-5050, USA}
\author{Sai Mu}
 \email{mus@mailbox.sc.edu}
 \affiliation{
 Department of Physics and Astronomy, University of South Carolina, Columbia, SC 29208, USA}
\author{Chris G. Van de Walle}
 \email{vandewalle@mrl.ucsb.edu}
 \affiliation{
 Materials Department, University of California Santa Barbara, Santa Barbara, CA 93106-5050, USA}
 
\date{\today}% It is always \today, today,
             %  but any date may be explicitly specified

\pacs{}% insert suggested PACS numbers in braces on next line

\maketitle %\maketitle must follow title, authors, abstract and \pacs

\subsection{Sensitivity of electro-optic coefficients to the choice of functional}
\label{sec:scissor}

We use the local density approximation (LDA) as the exchange-correlation functional in our density functional theory (DFT) calculations.
LDA underestimates the band gap of semiconductors and insulators; here we test to what extent this could affect our results for electro-optic (EO) coefficients.

EO coefficients have contributions from electronic, ionic, and piezoelectric responses.
According to Eqs.~(2) and (5) in the main text, each contribution includes at least one variable that could be affected by the band gap, including dielectric permittivity $\epsilon_{ij}$, nonlinear optical susceptibility $d_{ij}=\frac{1}{2}\chi^{(2)}_{ij}$, Raman susceptibility $\alpha_{ij}^m$, and elasto-optic coefficients $p_{ij}$.
To assess the impact of the band gap on these properties, we apply a scissor correction which adjusts the conduction-band energies such that the band gap matches the experimental value.

We use AlN and $\mathrm{Al}_{0.5}\mathrm{Sc}_{0.5}\mathrm{N}$ in primitive cells as examples, with LDA-calculated band gaps of 4.38 and 3.25 eV, respectively.
To adjust the gap to experimental values,~\cite{1994Guo_AlN_bandgap, 2019Baeumler_AlScN_gap} scissor corrections of 1.65 eV for AlN and 1.25 eV for $\mathrm{Al}_{0.5}\mathrm{Sc}_{0.5}\mathrm{N}$ are applied.
Table~\ref{scissor} presents a comparison of various properties calculated using LDA and the ``LDA + scissor correction'' approach.

In pure AlN, the ionic contribution $r^{\mathrm{ion}}_{13}$ dominates the total EO coefficient.
Adjusting the band gap reduces $r^{\mathrm{total}}_{13}$ by 10\% and $r^{\mathrm{total}}_{33}$ by 17\%.
%In $\mathrm{Al}_{0.5}\mathrm{Sc}_{0.5}\mathrm{N}$, $r^{\mathrm{total}}_{13}$ is enhanced more than $r^{\mathrm{total}}_{33}$.
In $\mathrm{Al}_{0.5}\mathrm{Sc}_{0.5}\mathrm{N}$, 
%the pure electronic part contributes most to $r^{\mathrm{total}}_{13}$.
the band-gap correction reduces $r_{13}^{\mathrm{elec}}$ and $r_{13}^{\mathrm{ion}}$, but increases $r_{13}^{\mathrm{piezo}}$, leading to a compensation in correction terms and reductions of $r^{\mathrm{total}}_{13}$ by only 7\% and $r^{\mathrm{total}}_{33}$ by only 11\%.
Our tests thus indicate that adjusting the band gap changes the total EO coefficients by less than  17\%.
We conclude that corrections to the EO coefficients due to the LDA underestimation of the band gap are relatively modest and will not affect our conclusions.

\newpage

\begin{table}[!h]
\centering
\caption{Comparison of various properties of AlN and $\mathrm{Al}_{0.5}\mathrm{Sc}_{0.5}\mathrm{N}$ calculated by LDA and with scissor correction. 
$\epsilon_{ij}$ and $\alpha_{ij}$ are dielectric perimittivity and Raman susceptibility.
$d_{ij}$ (pm/V), $p_{ij}$, $r_{ij}$ (pm/V) represent the nonlinear optical susceptibility, elasto-optic coefficients and EO coefficients in Voigt notation. The electronic, ionic and piezoelectric contributions of the EO coefficients are listed separately . 
}
\begin{tabular}{m{2cm}>{\centering} m{3cm}>{\centering} m{3cm}>{\centering} m{3cm}>{\centering\arraybackslash} m{3cm} }
  \toprule
  \multirow{2}*{} & \multicolumn{2}{c}{AlN} & \multicolumn{2}{c}{$\mathrm{Al}_{0.5}\mathrm{Sc}_{0.5}\mathrm{N}$} \\
  \cmidrule(lr){2-3}\cmidrule(lr){4-5}
  & LDA & Scissor & LDA & Scissor \\
  \midrule
  $\epsilon_{11}^{\rm elec}$ & 4.393 & 3.888 & 5.883 & 5.136 \\
  $\epsilon_{33}^{\rm elec}$ & 4.563 & 4.017 & 5.861 & 5.113 \\
  \hline
  $\alpha_{11}^{m}$ & -0.003 & -0.002 & 0.007 & 0.005 \\
  $\alpha_{33}^{m}$ & 0.006 & 0.004 & 0.003 & 0.002 \\
  \hline
  $d_{13}$ & -0.017 & 0.010 & -13.346 & -8.102 \\
  $d_{33}$ & -4.069 & -2.655 & -0.027 & 1.406 \\
  \hline
  $p_{11}$ & -0.069 & -0.059 & -0.012 & -0.015 \\
  $p_{12}$ & -0.055 & -0.043 & 0.011 & 0.022 \\
  $p_{13}$ & -0.083 & -0.075 & 0.014 & 0.024 \\
  $p_{31}$ & -0.026 & -0.017 & -0.021 & -0.015 \\
  $p_{33}$ & -0.120 & -0.104 & 0.028 & 0.038 \\
  $p_{55}$ & -0.062 & -0.068 & -0.075 & -0.088 \\
  \hline
  $r_{13}^{\rm elec}$ & 0.003 & -0.002 & 1.542 & 1.229 \\
  $r_{33}^{\rm elec}$ & 0.782 & 0.658 & 0.003 & -0.215 \\
  $r_{13}^{\rm ion}$ & -0.735 & -0.631 & 0.878 & 0.740 \\
  $r_{33}^{\rm ion}$ & 1.516 & 1.301 & 0.269 & 0.061 \\
  $r_{13}^{\rm piezo}$ & -0.193 & -0.197 & 0.529 & 0.780 \\
  $r_{33}^{\rm piezo}$ & -0.593 & -0.541 & 1.625 & 1.851 \\
  $r_{13}^{\rm total}$ & -0.925 & -0.830 & 2.949 & 2.749\\
  $r_{33}^{\rm total}$ & 1.705 & 1.418 & 1.897 & 1.697 \\
  \bottomrule
\end{tabular}
\label{scissor}
\end{table}

\newpage

\subsection{Lattice parameters of $\mathrm{Al}_{1-x}\mathrm{Sc}_x\mathrm{N}$ alloys}
\label{sec:lattice_geometry}
\begin{figure*}[h]
    \centering
    \includegraphics[scale=0.75]{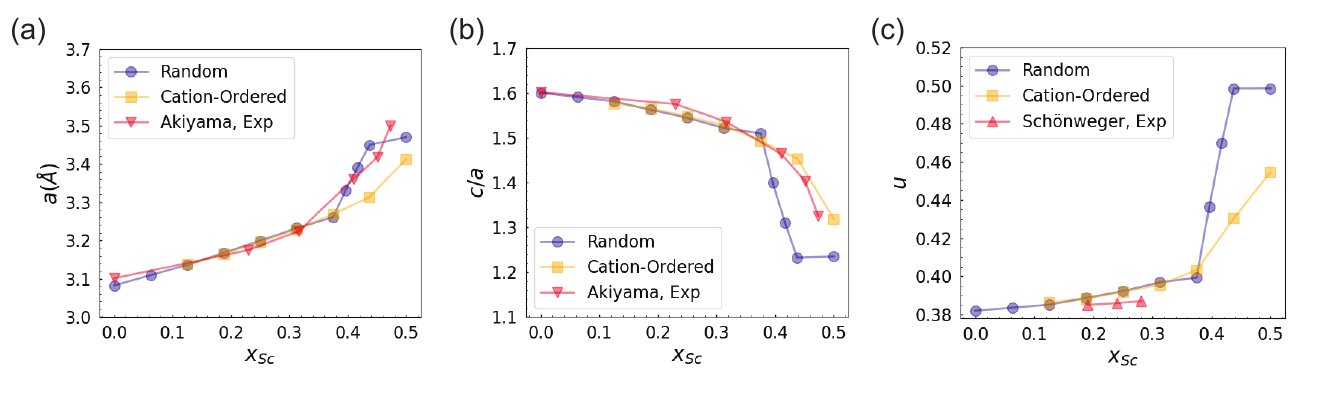}
    \caption{(a) Lattice parameter $a$, (b) $c/a$ ratio and (c) internal displacement parameter $u$ as a function of Sc concentration in $\mathrm{Al}_{1-x}\mathrm{Sc}_x\mathrm{N}$ alloys. Blue points represent averages over ten random configurations, while orange points represent averages over cation-ordered structures. Experimental values of $a$ and $c/a$ are from Akiyama {\it et al.},~\cite{Akiyama2009} while experimental $u$ values are from Sch\"{o}nweger {\it et al.}.~\cite{Schonweger2022} }
    %The experimental lattice parameter of GaN, indicated by the dotted line in panel (a), is from Ref.~\onlinecite{1996APL_lattice_GaN}.
    \label{fig_lattice}
\end{figure*}

Results for structural parameters of $\mathrm{Al}_{1-x}\mathrm{Sc}_x\mathrm{N}$ alloys are shown in Fig.~\ref{fig_lattice}.
As stated in the main text, we randomly generate ten configurations at each Sc concentration and compute the averaged lattice parameters, labeled as ``Random" in Fig.~\ref{fig_lattice}.
We also generate eight cation-ordered structures at each Sc concentration, again computing averages, labeled as ``Cation-Ordered".
Figure \ref{fig_lattice} illustrates that in random alloys the phase transition from wurtzite to layered-hexagonal occurs between 37.5\% and 43.75\% Sc concentration.
%, in fair agreement with previous studies as cited in our paper.
In contrast, the cation-ordered structures stay wurtzite even at 43.75\% and 50\% Sc concentrations.

\newpage

%\subsection{Spread of EO coefficients of $\mathrm{Al_{0.5}Sc_{0.5}N}$ }
%Besides lattice geometries, we also compare EO coefficients, both clamped and piezoelectric contributions, of random and ordered structures, see Fig. \ref{figS2}.

\subsection{Components of the EO tensor of random AlScN alloys}
\label{sec:r_11_22}
\begin{figure*}[h]
    \centering
    \includegraphics[scale=0.6]{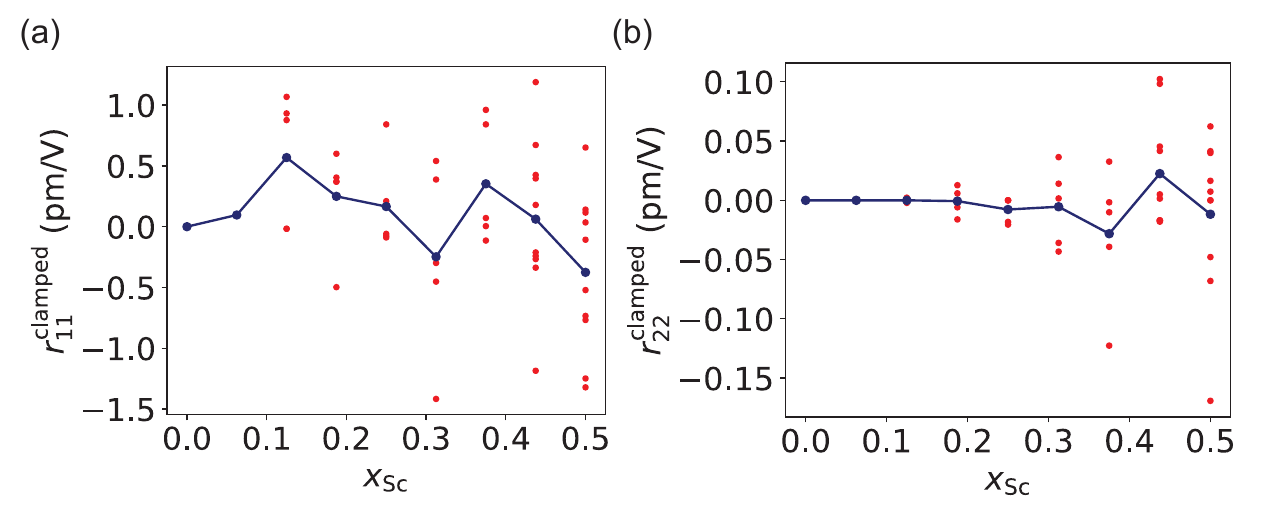}
    \caption{Clamped EO coefficients (a) $r_{11}^{\mathrm{clamped}}$ and (b) $r_{22}^{\mathrm{clamped}}$ of random AlScN alloys as a function of Sc concentration. The red points represent different configurations at each Sc concentration; the blue curve denotes averages over the multiple configurations at each Sc concentration. }
    \label{fig_SM_r11_r22}
\end{figure*}

Wurtzite AlN has $6mm$ point group symmetry, which specifies that there should be only three independent non-zero components in the EO tensor;
%, namely $r_{13}$, $r_{33}$ and $r_{51}$. The EO tensor 
in Voigt notation:
\begin{equation}
r_{hk} =\left(\begin{array}{ccc}
0 & 0 & r_{13} \\
0 & 0 & r_{13} \\
0 & 0 & r_{33} \\
0 & r_{51} & 0 \\
r_{51} & 0 & 0 \\
0 & 0 & 0
\end{array}\right)
\end{equation}
However, the supercells used for modeling AlScN alloys do not necessarily have the full wurtzite symmetry and can hence produce non-zero values for components such as $r_{11}$ and $r_{22}$.
In Fig.~\ref{fig_SM_r11_r22}, we plot the clamped EO coefficients $r_{11}^{\mathrm{clamped}}$ and $r_{22}^{\mathrm{clamped}}$. 
We observe that these values are non-zero, but in general they are small and tend to be zero when averaging over all configurations at each Sc concentration.
The variations of $r_{11}^{\mathrm{clamped}}$ are larger than those of $r_{22}^{\mathrm{clamped}}$. 
The difference can be attributed to the anisotropy of the structure, with the {\bf a} direction corresponding to ``armchair'' and the { \bf b} direction to ``zigzag'' configurations. 
Ignoring details of the alloy, the system has mirror symmetry for planes perpendicular to the {\bf b} axis (i.e., the \textit{a} plane of the wurtzite structure), but no such symmetry is present for planes perpendicular to the {\bf a} axis (the \textit{m} plane). 
The inherent mirror symmetry is responsible for the small values of $r_{22}^{\mathrm{clamped}}$.

\newpage

\subsection{Variations of $r_{33}$ among different configurations of random alloys}
\begin{figure*}[h]
    \centering
    \includegraphics[scale=0.7]{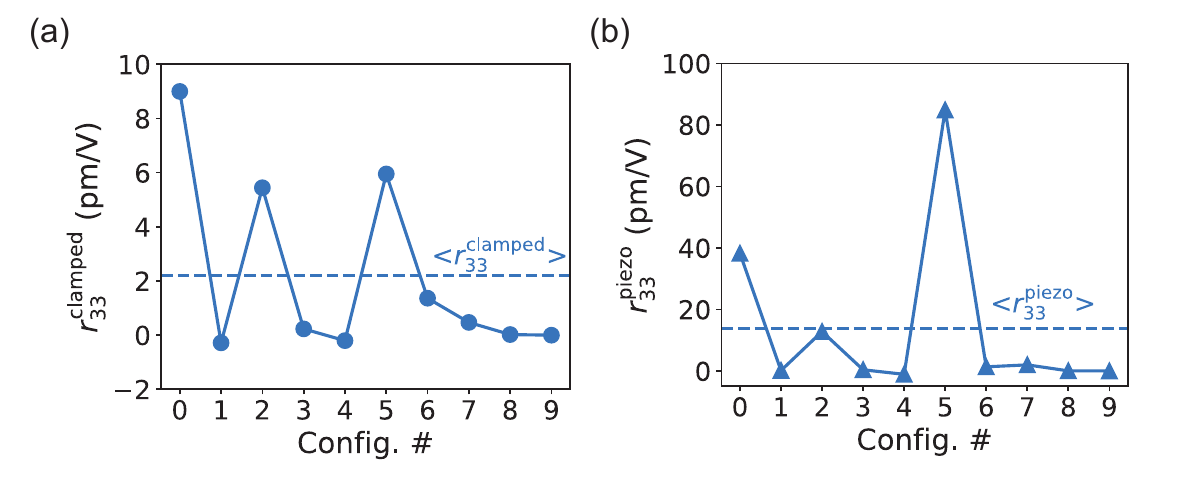}
    \caption{Clamped (a) and piezoelectric (b) EO coefficients of ten inequivalent configurations of $\mathrm{Al}_{0.5} \mathrm{Sc}_{0.5} \mathrm{N}$ random alloys. The dotted horizontal lines denote averaged values of ten configurations.}
    \label{fig_r33_variations}
\end{figure*}

%$\mathrm{Al}_{0.5} \mathrm{Sc}_{0.5} \mathrm{N}$}
\label{sec:variations_r33}
In $\mathrm{Al}_{1-x} \mathrm{Sc}_{x} \mathrm{N}$ random alloys, the variations of both clamped and piezoelectric EO coefficients among different configurations are small at low $x_{\mathrm{Sc}}$ but become significant at high $x_{\mathrm{Sc}}$. 
Figure \ref{fig_r33_variations} shows the large variations of $r_{33}^{\mathrm{clamped}}$ and $r_{33}^{\mathrm{piezo}}$ in random alloys at $x_{\mathrm{Sc}}=0.5$. 
Three out of ten configurations have large EO responses.
An analysis of pair distribution functions allowed us to identify a common structural feature of alloy configurations that give rise to high EO coefficients: the formation of same-species cation chains along the $c$ axis.

\newpage

\subsection{Bond strength in random and cation-ordered structures}
\label{sec:ICOHP_analysis}
The enhancements of piezoelectric EO coefficients in cation-ordered structures are attributed to the enhanced piezoelectric strain coefficients which further result from the large compliance constants. 
Large $S_{33}$ in cation-ordered structures reflects weak cation-nitrogen bonds along the $c$ axis. 
In order to evaluate the bond strength, crystal orbital Hamilton population (COHP) calculations\cite{COHP} are performed by employing the LOBSTER software package\cite{LOBSTER}, where the integrated COHP (ICOHP) up to the Fermi level serves as an indicator for the bond energy: the more negative values indicate stronger bonds.

\begin{figure*}[h]
    \centering
    \includegraphics[scale=0.65]{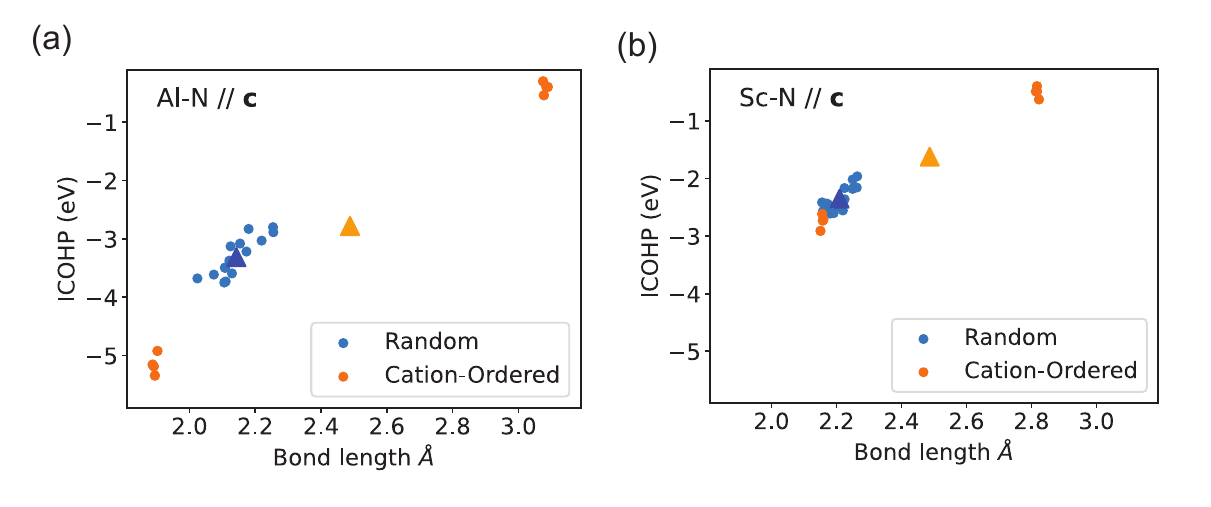}
    \caption{Integrated (over occupied states) crystal orbital Hamilton populations (ICOHP) for each cation–nitrogen interaction along the $c$ axis in 32-atom Random and Cation-ordered supercells of $\mathrm{Al_{0.5}Sc_{0.5}N}$ as a function of bond length: (a) Al–N bonds and (b) Sc–N bonds. 
    Averages are denoted using triangles. }
    \label{fig_ICOHP}
\end{figure*}

We analyze one structure from 32-atom random supercells and another from cation-ordered supercells of $\mathrm{Al}_{0.5}\mathrm{Sc}_{0.5}\mathrm{N}$. 
Figure \ref{fig_ICOHP} shows the bond energies of Al-N and Sc-N bonds along the $c$ axis as a function of bond length.
In cation-ordered structures, the magnitude of the averaged ICOHP for Sc–N and Al-N bonds becomes smaller than in random structures, indicating softer cation-nitrogen bonds along the $c$ axis.

In the case of Al-N bonds, weaker bonding means that the $u$ parameter increases (from its value $u$=0.381 in wurtzite AlN towards the higher value in a layered hexagonal structure).
Conversely, in the case of Sc-N bonds, softer bonds decrease $u$, towards the value in wurtzite.  
Both effects drive the system closer to the phase transition, which is
favorable for enhancing the EO response.

\newpage

\subsection{Independence of conclusions for strained structures on alloy configuration}
\label{sec:independence_strain}
\begin{figure*}[h]
    \centering
    \includegraphics[scale=0.6]{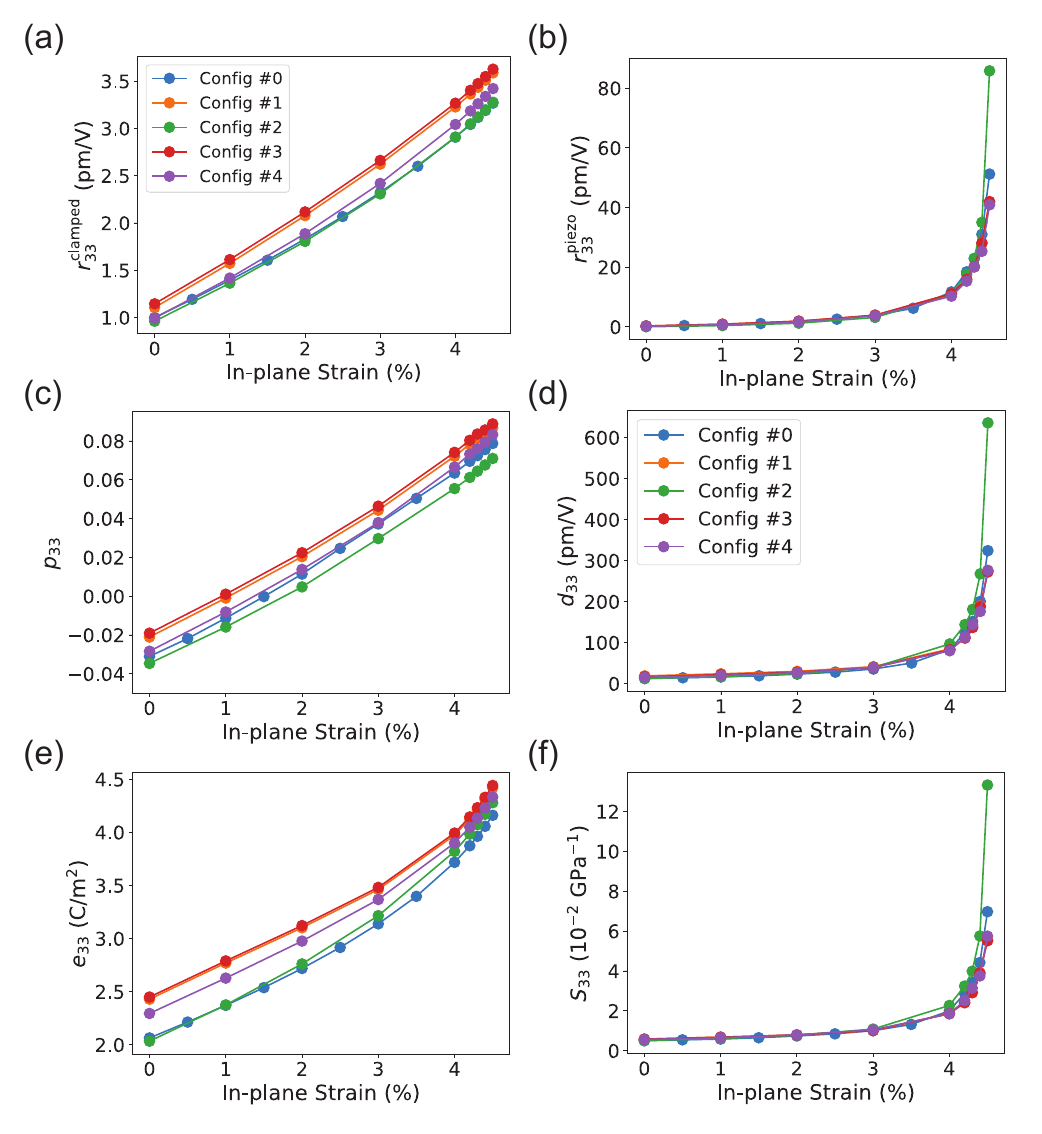}
    \caption{(a) Clamped EO coefficients $r^{\rm clamped}_{33}$, (b) piezoelectric EO coefficients $r^{\rm piezo}_{33}$, (c) elasto-optic coefficients $p_{33}$, (d) piezoelectric strain coefficients $d_{33}$, (e) piezoelectric stress coefficients $e_{33}$ and (f) compliance constants $S_{33}$ of $\mathrm{Al_{0.8125}Sc_{0.1875}N}$ as a function of in-plane tensile strain.}
    \label{fig_3Sc_strain}
\end{figure*}

Our investigations on the strain effect in the main text are based on one supercell at each Sc concentration of $\mathrm{Al}_{1-x} \mathrm{Sc}_x \mathrm{N}$ alloys. 
In Fig.~\ref{fig_3Sc_strain}, we investigate the EO coefficients of five inequivalent configurations of $\mathrm{Al_{0.8125} Sc_{0.1875} N}$ as a function of strain.
We observe almost identical enhancements in the clamped and piezoelectric EO coefficients as well is in their contributing factors ($p_{33}$, $d_{33}$, $e_{33}$ and $S_{33}$) for the different configurations. 
We also find that the critical strain is independent of the choice of configuration. 

\newpage

\subsection{Decomposition of $r_{33}$ in the presence of strain}
\label{sec:decomposition_r33}
\begin{figure*}[h]
    \centering
    \includegraphics[scale=0.8]{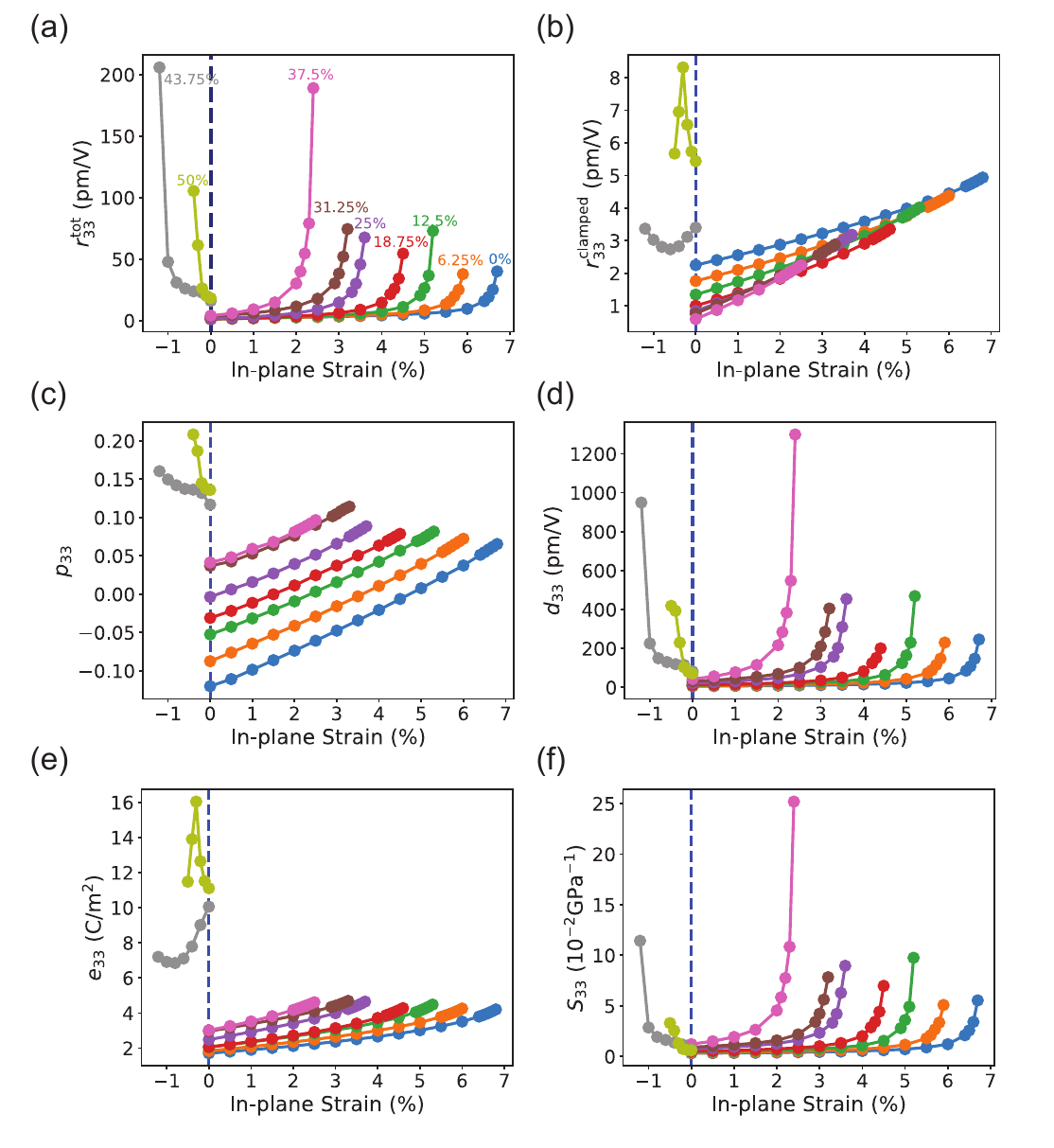}
    \caption{(a) Total EO coefficients $r^{\rm tot}_{33}$, (b) Clamped EO coefficients $r^{\rm clamped}_{33}$, (c) elasto-optic coefficients $p_{33}$, (d) piezoelectric strain coefficients $d_{33}$, (e) piezoelectric stress coefficients $e_{33}$ and (f) compliance constants $S_{33}$ of $\mathrm{Al}_{1-x}\mathrm{Sc}_x\mathrm{N}$ alloys as a function of in-plane strain.}
    \label{fig_strain_engi}
\end{figure*}

Figures~\ref{fig_strain_engi} (a) and (b) show the total and clamped EO coefficients of $\mathrm{Al}_{1-x}\mathrm{Sc}_x\mathrm{N}$ as a function of in-plane strain.
In contrast to the dramatic enhancement of $r^{\mathrm{piezo}}_{33}$, $r^{\mathrm{clamped}}_{33}$ increases almost linearly without exhibiting divergent behavior when $x_{\mathrm{Sc}} < 43.75\%$.
%At Sc concentrations of 43.75\% and 50\%, the non-monotonic behaviors of $r^{\mathrm{clamped}}_{33}$ are attributed to phase instability near the phase transition point.

Figure \ref{fig_strain_engi} (c)-(f) show the factors contributing to $r^{\mathrm{piezo}}_{33}$.
The piezoelectric coefficient $d_{33}$ and the compliance constant $S_{33}$ exhibit divergent behavior similar to $r^{\mathrm{piezo}}_{33}$.
A critical strain is present at each Sc concentration, decreasing as the Sc concentration increases.
As the in-plane strain approaches this critical value, the compliance constant $S_{33}$ increases dramatically, indicating that relatively small stresses can induce very large strain responses.
This softening of the material significantly enhances $d_{33}$, thereby amplifying the piezoelectric contributions to EO coefficients $r^{\mathrm{piezo}}_{33}$.

\begin{figure*}[h]
    \centering
    \includegraphics[scale=0.55]{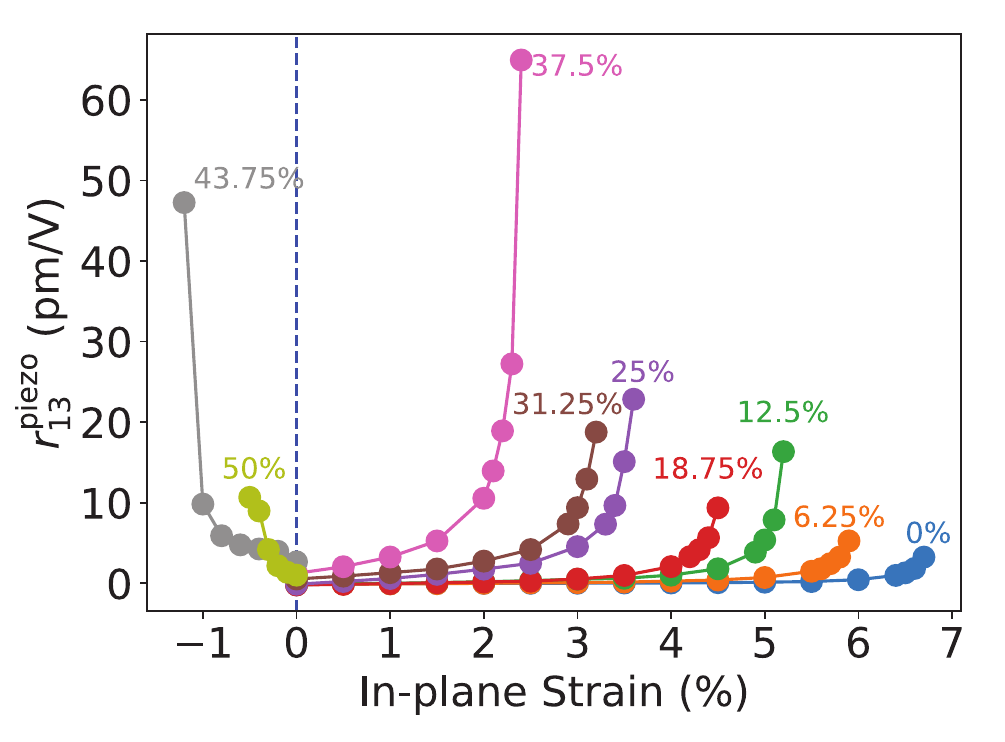}
    \caption{Piezoelectric contribution to the EO coefficient, 
    %$r_{13}$ of random and $c$-ordered AlScN alloys. (b) 
    $r_{13}^{\mathrm{piezo}}$, for random AlScN alloys as a function of in-plane strain.}
    \label{fig_r13}
\end{figure*}

% in addition to $S_{33}$, $d_{33}$, and $r^{\mathrm{piezo}}_{33}$, 
We note that $r^{\mathrm{piezo}}_{13}$ (and its components $d_{13}$ and $S_{13}$) exhibit similar behavior as a function of strain; see Fig.~\ref{fig_r13}.

\newpage

\subsection{Electro-optic coefficients of $a$-plane $\mathrm{(AlN)}_m/\mathrm{(ScN)}_n$ superlattices with $m+n=2, 4, 6$}
\label{sec:smaller_superlattice}
In the main text, we construct $a$-plane superlattices with $m+n=8$ at four different Sc concentrations, $\mathrm{(AlN)_7/(ScN)_1}$, $\mathrm{(AlN)_6/(ScN)_2}$, $\mathrm{(AlN)_5/(ScN)_3}$ and $\mathrm{(AlN)_4/(ScN)_4}$, and we observe large enhancements in both clamped and piezoelectric EO coefficients $r_{33}$ at 50\% Sc concentration ($\mathrm{(AlN)_4/(ScN)_4}$).

We have also generated three other $a$-plane
 superlattices, all at 50\% Sc concentration, namely $\mathrm{(AlN)_1/(ScN)_1}$, $\mathrm{(AlN)_2/(ScN)_2}$ and $\mathrm{(AlN)_3/(ScN)_3}$.
The clamped and piezoelectric EO coefficients $r_{13}$ and $r_{33}$ are shown in Table \ref{tab:smaller_superlattice}, together with the case of $m=n=4$. 
In all four $a$-plane superlattices, the total EO coefficients are enhanced compared to those in random alloys.
We observe that superlattices containing an odd number of layers of each material have larger EO coefficients than those containing an even number of layers.

\begin{table}[h]
\centering
\caption{Clamped and piezoelectric EO coefficients $r_{13}$ and $r_{33}$ (pm/V), in four $a$-plane $\mathrm{(AlN)}_m/\mathrm{(ScN)}_n$ superlattices with $m=n$.}
\label{tab:smaller_superlattice}
\begin{tabular}[t]{l>{\centering}p{0.1\linewidth}>{\centering}p{0.1\linewidth}>{\centering}p{0.1\linewidth}>{\centering\arraybackslash}p{0.1\linewidth}}
\toprule
$m=n=$ & 1 & 2 & 3 & 4 \\
\midrule
$r_{13}^{\mathrm{clamped}}$ & 0.03 & 2.24 & 0.58 & 0.48\\
$r_{13}^{\mathrm{piezo}}$ & 6.66 & 4.60 & 7.39 & 1.72\\
$r_{33}^{\mathrm{clamped}}$ & 4.23 & 3.42 & 8.28 & 14.92\\
$r_{33}^{\mathrm{piezo}}$ & 71.89 & 15.45 & 27.88 & 23.10 \\
\bottomrule
\end{tabular}
\end{table}%

\newpage

\subsection{Analysis of electro-optic coefficients of $a$-plane $\mathrm{(AlN)}_m/\mathrm{(ScN)}_n$ superlattices  with $m+n=8$}
\label{sec:superlattice}
In $a$-plane $\mathrm{(AlN)}_m/\mathrm{(ScN)}_n$ superlattices, both clamped and piezoelectric contributions to $r_{33}$ are enhanced with an increase of ScN layers.
At 50\% Sc concentration, the total EO coefficient $r_{33}$ goes up to 41 pm/V.
In contrast to what we observed in $\mathrm{Al}_{1-x}\mathrm{Sc}_x\mathrm{N}$ alloys, in $a$-plane superlattices clamped contributions rise with Sc concentration and are comparable to piezoelectric contributions even at high Sc concentrations.

\begin{figure*}[!h]
    \centering
    \includegraphics[width=0.8\linewidth]{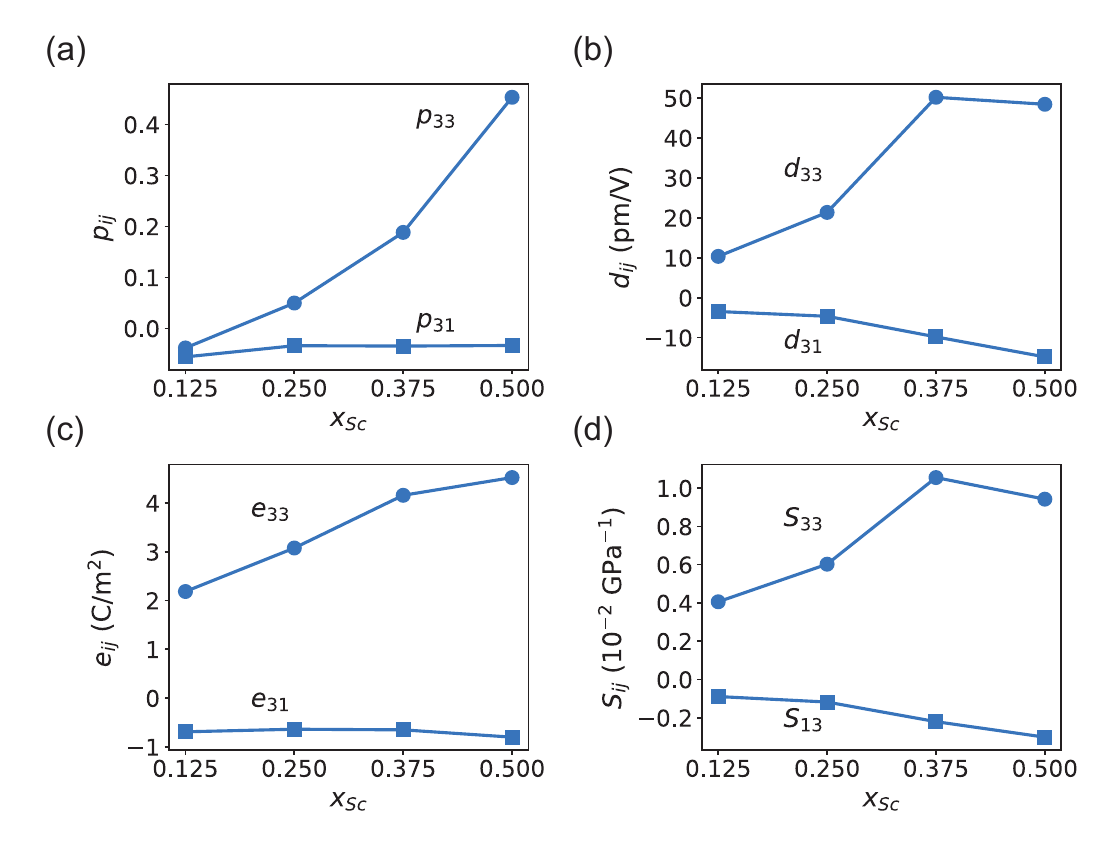}
    \caption{(a) Elasto-optic coefficients $p_{ij}$, (b) piezoelectric strain coefficients $d_{ij}$, (c) piezoelectric stress coefficients $e_{ij}$ and (d) compliance constants $S_{ij}$ of $a$-plane superlattices $\mathrm{(AlN)}_m/\mathrm{(ScN)}_n$ as a function of $x_{\rm Sc}=n/(m+n)$.}
    \label{fig_asl_piezo}
\end{figure*}

To explain why $a$-plane superlattices exhibit large piezoelectric EO coefficients at high Sc concentrations, the contributing terms to $r^{\mathrm{piezo}}_{33}$  and $r^{\mathrm{piezo}}_{13}$ are plotted in Fig.~\ref{fig_asl_piezo}.
According to Eqs.~(5) and (6) of the main text, the piezoelectric EO coefficients are derived from elasto-optic coefficients and piezoelectric strain coefficients, with the latter decomposed further into piezoelectric stress coefficients $e_{ik}$ and the compliance tensor $S_{kj}$.
From Fig.~\ref{fig_asl_piezo}(a), we observe that the elasto-optic coefficients $p_{33}$ significantly increase at higher Sc concentrations.
Additionally, the piezoelectric strain coefficients $d_{33}$ [Fig.~\ref{fig_asl_piezo}(b)] are notably high at 37.5\% and 50\% $x_{\rm Sc}$.
Combining these two factors, $a$-plane superlattices achieve large piezoelectric EO coefficients $r^{\mathrm{piezo}}_{33}$  at higher concentrations.

\begin{figure*}[h]
    \centering
    \includegraphics[scale=0.75]{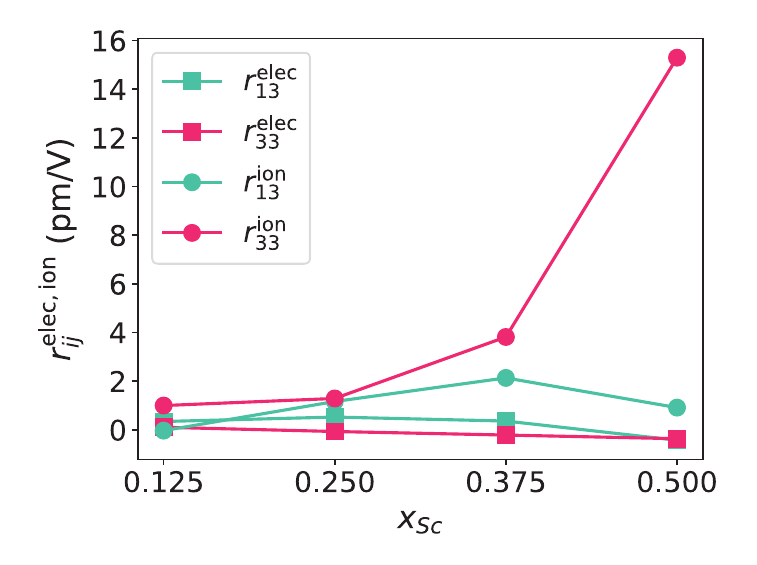}
    \caption{Ionic and electronic EO contributions $r_{33}$ and $r_{13}$ of $a$-plane superlattices $\mathrm{(AlN)}_m/\mathrm{(ScN)}_n$.}
    \label{fig_asl_el_ion}
\end{figure*}

The electronic and ionic contributions to the clamped EO coefficients $r_{33}^{\rm clamped}$ are plotted in Fig.~\ref{fig_asl_el_ion}.
The ionic response dominates the clamped EO coefficients, overshadowing electronic contributions, particularly at 50\% concentration.

\begin{figure*}[h]
    \centering
    \includegraphics[width=1\linewidth]{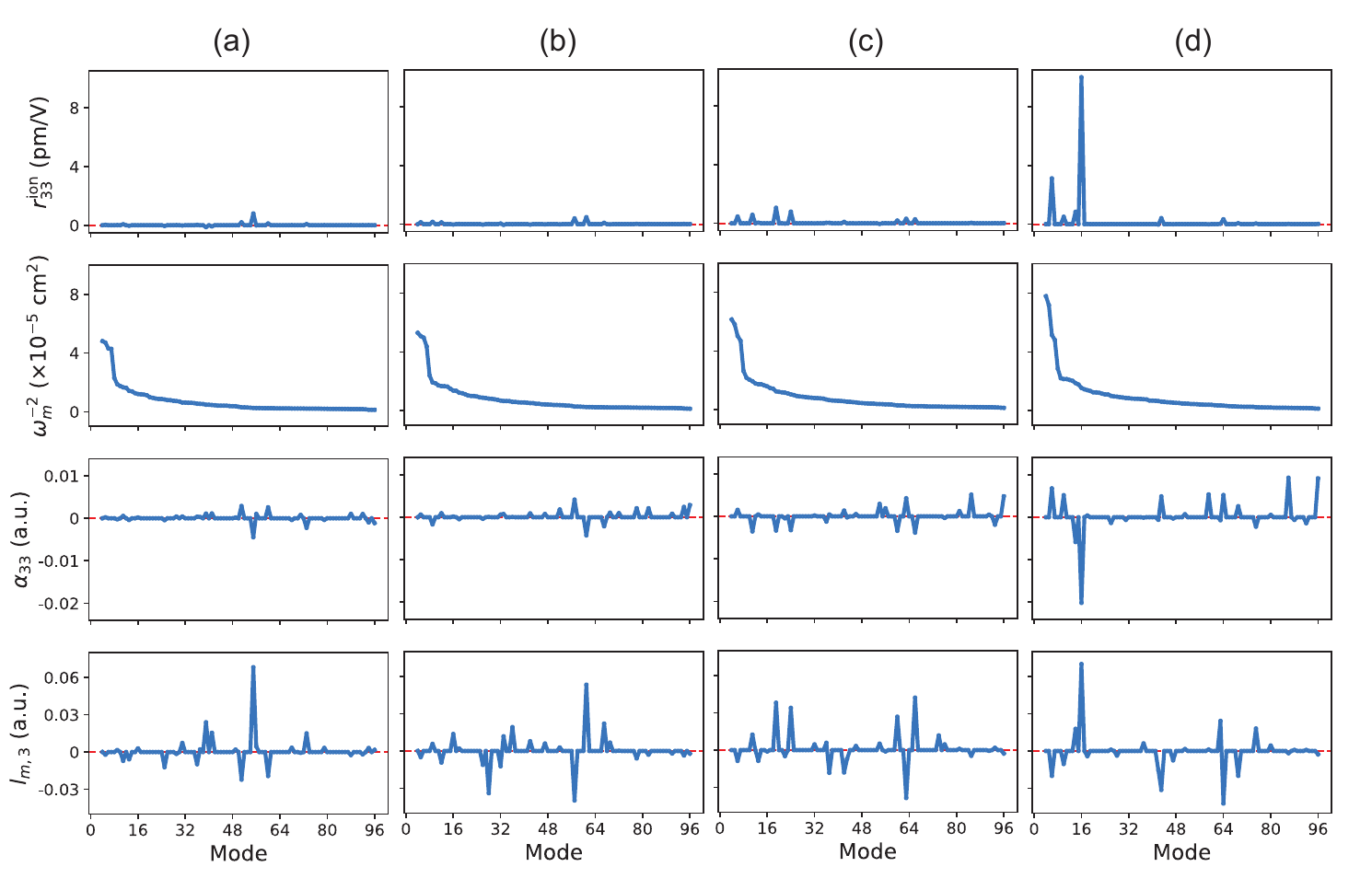}
    \caption{Mode-by-mode decompositions of $r^{\rm ion}_{33}$ for $a$-plane superlattices $\mathrm{(AlN)}_m/\mathrm{(ScN)}_n$ at (a) 12.5\%, (b) 25\%, (c) 37.5\% and (d) 50\% Sc concentrations. At each concentration, the panels show (from top to bottom) the ionic EO coefficient contributed by mode $m$, the frequency-dependent factor $\omega_m^{-2}$, the Raman susceptibility $\alpha^m_{33}$, and the mode polarity $l_{m,3}$.}
    \label{fig_mbm}
\end{figure*}

To elucidate the enhancement in the ionic contribution, we analyze $r^{\mathrm{ion}}_{33}$ through mode-by-mode decomposition in Fig.~\ref{fig_mbm}.
Given the use of 32-atom supercells, 96 phonon modes are included in the analysis.
%It was found that several modes strongly contribute to mode polarity $p_{m,3}$ across all four Sc concentrations.
%However, Raman susceptibilities $\alpha_{33}$ are negligible at Sc concentrations from 12.5\% to 37.5\%, leading to weak ionic EO response.
At 50\%, two shearing phonon modes, in which AlN and ScN layers move along $c$ in opposite directions, significantly contribute to the ionic response: mode 6 with relatively small Raman susceptibility and mode polarity but very low phonon frequency (139 cm$^{-1}$); and mode 16 with large Raman susceptibility and mode polarity but slightly higher phonon frequency (254 cm$^{-1}$).

\clearpage
\bibliography{EO_reference}